\documentclass{aa}  
\usepackage{natbib}
\bibpunct{(}{)}{;}{a}{}{,} 
\usepackage{graphicx}

\usepackage{txfonts}
\usepackage[breaklinks=true]{hyperref}
\hypersetup{colorlinks,linkcolor=red,citecolor=blue,urlcolor=blue}

\newcommand{\ubv}{\ensuremath{UB\,V}}
\newcommand{\bv}{\ensuremath{B\,V}}
\newcommand{\jks}{\ensuremath{(J-K_\mathrm s})}
\newcommand{\hks}{\ensuremath{(H-K_\mathrm s})}
\newcommand{\ks}{\ensuremath{K_\mathrm s}}
\newcommand{\jhks}{\ensuremath{JHK_\mathrm s}}
\newcommand{\logt}{log\,\textit{t}}
\newcommand{\ejk}{\ensuremath{E(J-K_{\mathrm s})}}
\newcommand{\ejh}{\ensuremath{E(J-H)}}
\newcommand{\ebv}{\ensuremath{E(B-V)}}

\begin{document}

   \title{A comparative study on the reliability of open cluster parameters}


   \author{M. Netopil
          \inst{1}
          \and
          E. Paunzen\inst{1}
          \and
          G. Carraro\inst{2,3}
          }

   \institute{Department of Theoretical Physics and Astrophysics, Masaryk University, Kotl\'a\v{r}sk\'a 2, 611 37 Brno, Czech Republic \\
             \email{mnetopil@physics.muni.cz}
         \and
         European Southern Observatory, Alonso de Cordova 3107, Casilla 19001, Santiago 19, Chile  
         \and
         Dipartimento di Fisica e Astronomia Galileo Galilei, Universit\`a di Padova, Vicolo Osservatorio 3, I-35122, Padova, Italy       
             }

   \date{Received <date>; accepted <date>}

 
  \abstract
   {Open clusters are known as excellent tracers of the structure and chemical evolution of the Galactic disk, however, the accuracy and reliability of  open cluster parameters is poorly known.}
   {In recent years, several studies aimed to present homogeneous open cluster parameter compilations, which are based on some different approaches and photometric data. These catalogues are excellent sources to facilitate testing of the actual accuracy of open cluster parameters.}
   {We compare seven cluster parameter compilations statistically and with an external sample, which  comprises the mean results of individual studies. Furthermore, we selected the objects IC~4651, NGC~2158, NGC~2383, NGC~2489, NGC~2627, NGC~6603, and  Trumpler~14, with the main aim to highlight differences in the fitting solutions.   }
   {We derived correction terms for each cluster parameter, using the external calibration sample. Most results by the compilations are reasonable scaled, but there are trends or constant offsets of different degree. We also identified one data set, which appears too erroneous to allow adjustments. After the correction, the mean intrinsic errors amount to about 0.2\,dex for the age, 0.08\,mag for the reddening, and 0.35\,mag for the distance modulus. However, there is no study that characterises the cluster morphologies of all test cases in a correct and consistent manner. Furthermore, we found that the largest compilations probably include at least 20 percent of problematic objects, for which the parameters differ significantly. These could be among others doubtful or unlikely open clusters that do not facilitate an unambiguous fitting solution.  }
   {}


   \maketitle
%

\section{Introduction}

Open star clusters are traditionally used as probes of the structure
\citep{Janes82} and chemical evolution \citep{Magr09} of the
Galactic disk.
 These groups of stars are particularly suitable tracers for two main reasons.
 First, they are spread all over the disk and span the entire disk
lifetime. Second, their fundamental properties (age, distance,
metallicity, and reddening) can be derived in a statistical manner, which,
 in principle, is more solid than
for individual stars if one assumes that all cluster stars are born at the
same time and in the same Galactic volume.

In the last couple of decades, several photometric surveys or catalogues  significantly increased the amount of data, and the number of open star clusters (or candidates) in
the Milky Way: for example, ASCC--2.5 \citep{Khar01}, SDSS \citep{Alam15}, 2MASS \citep{Skru06}, or VVV \citep{Saito12}, to mention just a few of them. 
This dramatically changed the approach of the study of these objects,
which can only be conducted on an individual cluster basis  for a few
particularly relevant objects.

As a consequence, several compilations of parameters started to appear in
the literature, which were derived using some semi-automatic tools. Some of these compilations include parameters for hundreds or even thousands of open clusters. It is obvious that the time spent on each object is quite limited. This also affects the usual careful  by-eye inspection  and consultation of the literature.  Therefore, a proper comparison of the methods and results is of importance.

These methods typically start from the distribution of stars in the
colour magnitude diagram (CMD) and fit with isochrones to simultaneously
derive age, distance, and reddening. No effort is normally paid to explore
the metallicity space, however, most of the open clusters are, within the errors, solar abundant \citep[see e.g.][]{heiter14}. More sophisticated methods add on kinematics, some statistical cleaning of
the CMD, and an analysis of the  spatial
distribution of the stars. The degree of the automatisation is quite different, although a visual inspection of the results and some manual interaction are always included. There is no work yet with a large open cluster sample that relies completely on automatic pipelines. To our knowledge, \citet{Perren15}  first  presented a suite of tools that allow a fully automated analysis, but so far  it has been applied to a limited number of objects.

While the potentials of these homogeneous compilations are obvious,
difficulties arise when inferred parameters for a given cluster from
different methods or data sources are compared and found very discrepant, thus mining the
reliability of the various underlying techniques.

In this paper, we perform a comparison of seven of these compilations with the main aim to highlight differences in the fitting
solutions.  Additionally, we present a few test cases (\object{IC 4651}, \object{NGC 2158}, \object{NGC 2383}, \object{NGC 2489}, \object{NGC 2627}, \object{NGC 6603}, and  \object{Trumpler 14})
to show how the assistance of an even quick by-eye inspection and a proper
consultation of the literature can help enormously to avoid unpleasant
mistakes. Furthermore, we compare the compilations with results from individual studies to derive correction terms and the errors for all parameters.

The paper is organised as follows: Sect. \ref{sect:compilations} presents an overview of studies that provide large sets of open cluster parameters. In Sect. \ref{sect:global}, we compare the results of the studies to the largest compilation and present some basic statistics. In Sect. \ref{sect:individual}, we discuss the results for some selected open clusters on the basis of the CMDs, and discuss global differences in the isochrone fitting procedure. Section \ref{sect:recal} compares all studies with the mean results of individual studies, and presents correction terms for the parameters and an error analysis. Finally, Sect. \ref{sect:conclusion} concludes the paper.

\section{The data compilation}
\label{sect:compilations}
Several works have investigated a large number of open clusters in a homogeneous way, at least in respect  to the interpretation of the CMD. However, different data sources were used, either photometry in the visual compiled from different papers or all-sky, near-infrared (NIR) 2MASS measurements \citep{Skru06}. 

In the following, we discuss the approaches of the considered studies (also named as surveys later on). Furthermore, we present an overview in Table \ref{tab:systcomp}, listing the size of the respective samples,   photometric systems, isochrone sets, metallicity, and extinction ratios.    

The analysis by \citet[][hereafter B11]{Buko11} is based only on 2MASS, $J$ versus \jks\ diagrams. A statistical cleaning procedure
was applied using the information of stars in the vicinity of the cluster area as described in \citet{Bica05}. Stars that were found to be outliers from the main sequence (MS) and
red giant clump were removed by hand after visual inspection. The distance modulus, reddening, and age were derived by fitting isochrones, which were shifted in both directions in the CMD with a step of 0.01\,mag. The solution giving the smallest $\chi^{2}$ 
was taken as the final solution.

\citet{Glus10} and \citet{Kopo08}, hereafter G10, used the data from 2MASS and for some fainter objects the data from the UKIDSS Galactic Plane Survey \citep{Luca08} to derive $J$ versus $(J-H)$ and \ks\ versus \jks\ diagrams. Beside the CMDs, the Hess-diagrams, which  also represented the spatial density of stars in the CMD, were applied. They performed a search for new open clusters in the Galactic disk ($\left| b \right|$\,$<$\,24\degr) and determined their parameters. 
Their sample is biased towards old ages, larger reddening values,  and distances larger than 1\,kpc from the Sun.

We merged the cluster parameters published by
\citet[][hereafter K05]{Khar05a,Khar05b}, resulting in 650 open clusters. The estimation of cluster parameters is based on the ASCC--2.5 catalogue \citep{Khar01}. It includes proper motions and Tycho-2 photometry, which is
transformed to standard Johnson \bv\ magnitudes. The catalogue is limited to $V$\,$<$\,14\,mag with a 90 percent completeness at $V$\,$\approx$\,11.5\,mag \citep{Hog00}. K05 concentrated on post-MS isochrones because the CMDs of many clusters in their sample  
present the evolved portions of the upper MS, and the pre-MS, observed at relatively faint absolute magnitudes, should not been visible in the cluster diagrams.
As a first step, the distance and reddening were 
evaluated  using values primarily from the literature, most of them from \citet[][and private communication]{Lokt01}. For about 200 clusters,
they determined or revised cluster distances and reddening on the basis of supplementary data on spectral classes of the most probable members
available from the ASCC--2.5 and the Tycho-2 Spectral Type Catalog \citep{Wrig03}. The standard relations of \citet{Schm82} and \citet{Stra92} were used
to fit the MS. The individual age of the stars was then derived from their locations in the CMD with respect to the isochrone grid. The final
cluster age is the mean of all individual values.

\citet[][hereafter K13]{Khar13} derived the cluster parameters using 2MASS photometry and the kinematic data of the PPMXL catalogue \citep{Roes10}. The method is
based on that of K05, but for the \jhks\ domain. The  description is presented in \citet{Khar12}. They introduced quality criteria for the photometry and proper motion data.
The interstellar reddening was determined with the help of a \hks\ versus $(J-H)$ diagram and
checked with the $Q$-method for 2MASS photometry. For the determination of the cluster distance, isochrones were fitted to the \ks\ versus $(J-H)$ and \ks\ versus 
\jks\ diagrams. They found a shift for $(J-H), $ which was typically of about a few hundreds of magnitude varying from cluster to cluster. It
was corrected by an empirical correction $\Delta H$ for each cluster. The ages of older clusters were determined by averaging the ages of the turn-off (TO) stars.
The age determination of younger clusters was carried out with the isochrone fitting technique. The authors note that the pipeline was run by and under control of
a single team member, the first author, to keep the results as homogeneous as possible. They provide cluster parameters for 3006 objects in total, also including  young associations or globular clusters. Besides the membership probabilities derived from proper motion data, they also list membership values for each star based on  photometric colours. 

\citet[][hereafter L01]{Lokt01} used data in several photometric systems to derive the cluster parameters. Each photometric system was
treated separately, and the final values were weighted and averaged. The weighting  depends on the photometric system and on the number of available
measurements and data quality. For the \ubv\ data, the reddening was determined using the $Q$-method
\citep{John53} and  standard relations by \citet{Khol81}. The distance and  age were derived by fitting isochrones taken from \citet{Bert94}. The choice of
metallicity is not listed, but it is most likely the solar value Z = 0.020. For the \bv\ data, all three free cluster parameters were determined simultaneously by fitting isochrones. From $DDO$ photometry, only 
the reddening and distance were determined using the method and standard relations from \citet{Jane75}, valid for G and K-type giants.
For the $uvby\beta$ system, isochrones were transformed from the \ubv\ system and then fitted. For the metallicity,
nearly solar is given, but without any specific value. For the $RGU$ system, again, the $Q$-method and the standard relations from \citet{Stei68} were applied.
The final mean distance moduli were shifted by 0.153\,mag to bring the distance scale in accordance with the adopted Hyades distance modulus of 3.42\,mag. For our analysis, we have
corrected for this shift.

\citet[][hereafter T02]{Tadr02} updated the cluster parameters published by \citet{Tadr01}. They were using Johnson \ubv\ CCD measurements for 160 open clusters (55 of them have
only \bv\ data) taken from the literature until 2000. The detailed list of references for each cluster is listed in \citet[][ Table 1]{Tadr01}. For the determination
of the reddening and distance modulus, the standard relations by \citet{Schm82} were taken. Evolved stars (neither the corresponding spectral types nor luminosity classes are listed) 
were excluded from the fitting procedure. The effect of differential reddening was taken into account using the
method of \citet{Burk75}. Finally, the age of the cluster was estimated using isochrones. 

There is an extensive series of papers \citep{Tadr08b, Tadr08a, Tadr09a, Tadr09b, Tadr11, Tadr12a, Tadr10, Tadr12b} by the working group of A.L. Tadross (hereafter T08) using 
2MASS, \jhks\ photometry to derive cluster parameters mainly of, at that time, unstudied aggregates. They fitted isochrones to $J$ versus $(J-H)$ and \ks\ versus \jks\ diagrams. The pre-selection of stars in the
cluster area on the basis of  the observational uncertainties,  membership criteria, and  used values of the total-to-selective extinction ratio ($R_V$), is not the same for all of the papers. For example, \citet{Tadr09b} adopted $R_V$\,=\,3.1, and \citet{Tadr09a} lists $R_V$\,=\,3.2. The paper series concentrated on the analysis of previously unstudied open clusters and their sample is biased towards distances larger than 1\,kpc from the Sun.


\begin{table*}
\caption{Overview of the surveys and the adopted parameters.} 
\label{tab:systcomp} 
\centering 
\begin{tabular}{l l c l c } 
\hline\hline 
Ref. & System & No. OC's & Z / Isochrone set / Group & $R_V$ and extinction ratios\\ 
\hline 
B11 & \jhks & 754 & 0.019 / \citet{Gira02} / P & $R_V = 3.1$; $A_J = 0.276 A_V$; $E(J-K_{\mathrm s}) = 0.52 E(B-V)$ \\
G10 & \jhks & 168 & 0.019 / \citet{Gira02} / P & $R_V = 3.1$; $A_J = 0.276 A_V$; $E(J-H)=0.33E(B-V)$\\
K05 & \bv & 650 & 0.019 / \citet{Gira02} / P & $R_V = 3.1$ \\
K13 & \jhks & 2808 & 0.019 / \citet{Mari08} / P & $R_V = 3.1$; $A_{K_{\mathrm s}} = 0.67\,E(J-K_{\mathrm s})$; $E(J-K_{\mathrm s}) = 0.48 E(B-V)$\\
L01\tablefootmark{a} & visual & 424 & 0.020 / \citet{Bert94} / P & $R_V = 3.34$\\
T02 & \ubv & 160 & 0.020 / \citet{Meyn93} / G & $R_V = 3.25$\\
T08 & \jhks     & 282 & 0.019 / \citet{Mari08} / P & $R_V = 3.1/3.2$; $A_J = 0.276 A_V$; $E(J-K_{\mathrm s}) = 0.488 E(B-V)$        \\
\hline 
\end{tabular}
\tablefoot{The isochrone sets are either from the Padova (P) or Geneva (G) group. 
\tablefoottext{a}{L01 used various photometric systems in the visual range, see discussion in Sect. \ref{sect:compilations}. We list here the isochrone set used by them for the \ubv\ data.}

}
\end{table*}


Table \ref{tab:systcomp} shows that most studies rely on similar isochrone sets and adopted a comparable (solar) metallicity. We note that the isochrones by \citet{Gira02} and \citet{Mari08} are identical except for the asymptotic giant branch phase. Differences in the results owed to isochrones are not expected among studies in the NIR because in the isochrone fitting procedure, in general,
more weight is given to earlier evolutionary phases, such as the TO point or
red giant clump. However,  studies in the visual (K05, L01, and T02) are possibly affected because isochrones by two groups are involved: Padova (P) and Geneva (G).  

\begin{figure}
\resizebox{\hsize}{!}{\includegraphics{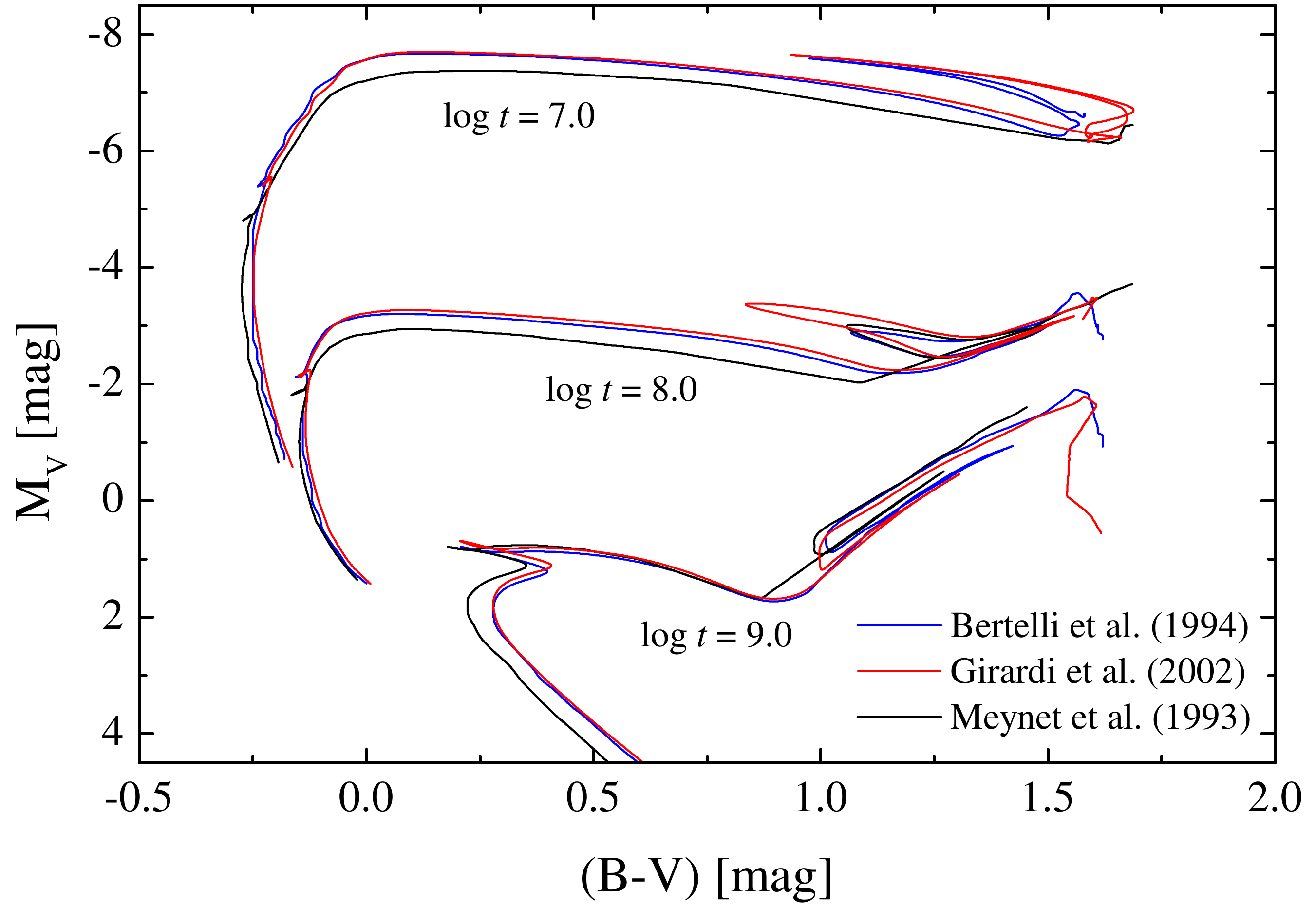}}
\caption{Comparison of isochrones used by studies in the visual.} 
\label{fig:isocomp}
\end{figure}

Figure \ref{fig:isocomp} shows the isochrones by \citet{Bert94}, \citet{Gira02}, and \citet{Meyn93} in the $(B-V)/V$ plane for some selected ages. We recall that L01 made use of several photometric systems, and the individual results are combined by  weight. Therefore, the influence by a single system (or a specific data set) can be hardly reproduced. However, \ubv\ is certainly still the dominating photometric system in open cluster research, and for many clusters probably the only available source. The studies by K05 and T02 used isochrones only for the determination of age, while distance and reddening were obtained with other methods. Therefore,  for the sake of  better visibility, the isochrones are only shown down to few magnitudes fainter than the TO point. In the following, we discuss the possible influence on the determination of age by a visual comparison, if no other shifts in colour or absolute magnitude were applied. The first two isochrone sets (both by the Padova group) agree very well for most evolutionary phases with differences of about 0.01\,mag for $(B-V)$ or about 0.1\,mag at maximum for the absolute magnitude $M_V$. These differences are negligible compared to the typical scatter in the CMD of open clusters (see e.g. Fig. \ref{fig:ic4651}, \ref{cmd1}, or \ref{cmd2}). Larger differences are apparent along the red giant loop, in particular, at an age of \logt\ = 8.0\,dex. However, the determination of possible systematic differences of the resulting age is virtually impossible. It strongly depends on the number of cluster stars, their distribution in the CMD, and the weight one assigns to the position of individual red giants in the CMD compared to the position of TO stars.

The red giant loop in the isochrones by \citet{Meyn93} agree well with \citet{Bert94}, but there is a mismatch of the absolute magnitudes for the blue hook and the sub-giant branch. To reach a reasonable  agreement between these branches, a younger age of about 0.1\,dex has to be adopted from the isochrone set by \citet{Meyn93}. This difference vanishes at an age of \logt\ $\sim$ 8.5\,dex. The 1\,Gyr isochrone by \citet{Meyn93} would match the others very well by applying a shift of 0.05\,mag to redder colours. Bringing the isochrones into agreement is more difficult when choosing a different age, but if considering the bluest turn-off colour one could argue that an $\sim$0.05\,dex older isochrone from the set by \citet{Meyn93} is necessary. We therefore conclude that there should be no significant difference in the results of the age among the open cluster studies in the visual simply owing to different isochrone sets. 


\section{Global differences between the studies}
\label{sect:global}


\begin{figure*}
\centering
\includegraphics[width=170mm]{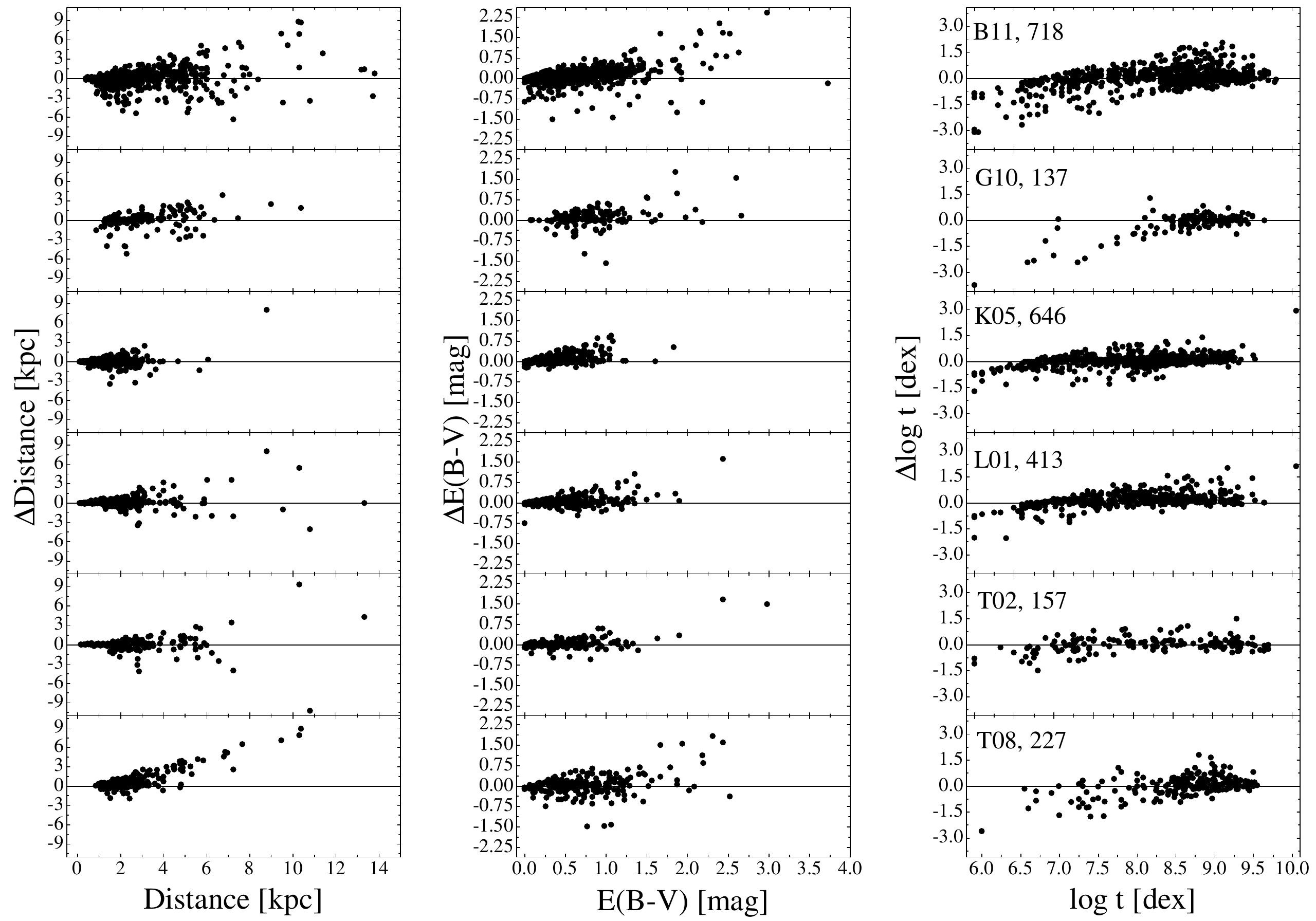}
\caption{Comparison of the star cluster parameters for objects in common with K13. The differences are calculated
in the sense `K13 -- reference'. In the right-most column, the designation of the individual references (see text) and the number of clusters in common
are listed.} 
\label{fig:statistics}
\end{figure*}



\begin{table*}
\caption{Basic statistical properties of the data sets as shown in Fig. \ref{fig:statistics}}
\label{tab:distributions}
\centering
\begin{tabular}{cccccccccccc}
\hline
\hline 
Reference & \multicolumn{3}{c}{$\Delta$ Distance} & \multicolumn{3}{c}{$\Delta$ Reddening} & \multicolumn{3}{c}{$\Delta$ Age} & N \tablefootmark{a} & AF \tablefootmark{b}\\
          & \multicolumn{3}{c}{[kpc]} & \multicolumn{3}{c}{[mag]} & \multicolumn{3}{c}{[dex]} & & \\
          & Mean & Mode & SD & Mean & Mode & SD  & Mean & Mode & SD  \\
\hline
B11     & $-$0.025 & $-$0.196 & 1.411   & +0.080 & +0.120  & 0.319      &       +0.023  & 0 & 0.615        & 718  & 51\\
G10     & +0.080 & 0 & 1.358    & +0.084 & $-$0.001  & 0.367    &       $-$0.172        & 0 & 0.650        & 137  & 50\\
K05     & +0.004 & 0 & 0.525    & +0.057 & 0  & 0.129   &       +0.041  & 0 & 0.345        & 646  & 78\\
L01     & +0.068 & 0 & 0.795    & +0.042 & 0  & 0.164   &       +0.163  & $-$0.177 & 0.434         & 413  & 71\\
T02     & $-$0.059 & $-$0.362 & 1.496   & +0.043 & $-$0.020  & 0.232    &       +0.002  & 0 & 0.422        & 157  & 68\\
T08     & +0.833 & 0 & 1.469    & +0.028 & 0  & 0.384   &       +0.009  & +0.001 & 0.588   & 227  & 38\\
\hline
\end{tabular}
\tablefoot{
\tablefoottext{a}{The number of objects in common with K13.}\tablefoottext{b}{The agreement factor (AF) is the percentage of objects in common for those all parameters are within specific limits.}}
\end{table*}


Figure \ref{fig:statistics} shows the differences of the distance, reddening, and age of the investigated references. The differences are calculated
in the sense `K13 -- reference' because K13 presents by far the largest open cluster sample and includes most of the objects studied by the other groups. All diagrams are on the same scale to enable an immediate comparison. We performed a statistical analysis of
the distributions of the data as plotted in Figure \ref{fig:statistics}. To accomplish this, we used the mean, mode, and  standard deviation (SD). A comparison of the individual parameters alone is not sufficient. Usually, follow-up works use a complete parameter set or a combination, the distance and the reddening, for example, to derive  the luminosity function of a cluster. The three cluster parameters (four if also considering metallicity) are strongly coupled in the isochrone fitting procedure. Thus, a wrong choice of one parameter affects all others as well. We therefore need a measure of the agreement in the three-dimensional space, which we call the agreement factor (AF). To accomplish this, we use the percentage of the clusters in common for those  parameters that are within specific limits. We adopt the mean standard deviations as limit: 0.25\,mag for the reddening \ebv, 0.8\,mag for the true distance modulus, and 0.5\,dex for \logt. Table \ref{tab:distributions} lists all parameters and  derived values.

{\it Distance}: There is a clear trend with the distance visible for T08. The most distant clusters in K13 are significantly less distant in T08, which is also
apparent in the mean value of the distribution. This is very probably because of the different approaches to interpreting the 2MASS CMD (see discussion in Sect. \ref{sect:diffiso}). A similar trend might be also indicated in the data sets of B11 and G10. However, most of the objects with strong deviating results were not studied so far on an individual basis in the visual, and there is also no single data entry in WEBDA. Thus, based on current knowledge, it is difficult to decide which distance scale is correct. All studies, but K05 and L01, show large standard deviations of almost 1.5\,kpc. Therefore, the current uncertainties of the distance for open clusters do not allow us to use them for tracing the global
characteristics of the Milky Way.  

{\it Reddening}: The mean of all six references is positive, which means that K13 overestimate the reddening in a statistical sense. 
The mean of the standard deviation of about 0.3\,mag converts to almost one magnitude for the total absorption $A_{\mathrm V}$. 

{\it Age}: For all references, there is a clear trend for the differences in age in the way that many young open clusters from K13 are significantly
older in the other sources. For G10 and T08 (both included new and/or unstudied clusters), this trend even goes up to $\log t$\,=\,8.5\,dex.
This fact might be explained by (not) taking into account apparent giants as members of the respective clusters. For one object (NGC~2489, see Sect. \ref{sect:individual}), we noticed that the study by K13 has not considered  red giants, although a high kinematic membership probability was derived for the stars.  This could be an accidental example and does not allow for an overall conclusion. 

A special
case is the data set of B11. There is a linear dependency over the complete age range with a bandwidth of about 2.5\,dex. Although B11 and K13 use
2MASS photometry, the cause of this effect is not obvious. The large spread over the whole age range lead us to 
conclude that we are currently still in the same situation as described in \cite{Merm81}, who defined 14 age classes rather than trying to
derive an exact age for open clusters.

{\it Agreement factor}: This value is a measure of the overall agreement between different studies and methods, but does not imply that the results also correctly reproduce the cluster morphology. The compilations by K05 and L01 agree well with K13 for more than 70 percent of the objects. However, as discussed in Sect. \ref{sect:diffiso}, the starting values by K13 were adopted from these references and have not changed much for numerous targets (see e.g. NGC~2383 in Fig. \ref{cmd1} and Table \ref{tab:paracomp}). The more independent studies, which  are based on 2MASS data as well (B11, G10, and T08), show a lower agreement level of about 50 percent or less. However, the samples by G10 and T08 consist mainly of more distant and previously unstudied objects. These clusters might show a  field star contamination that is too strong, complicating a correct isochrone fitting. The only independent study in the visual by T02 shows a higher agreement of almost 70 percent. This sample includes a large number of well-confirmed closer open clusters. These objects probably also show  a clear recognizable cluster sequence in the NIR.

Numerous objects that show quite large differences for  age and distance might be spurious open clusters as well. Targets, such as Antalova~2, Juchert~3, or  Miller~1, are only included in the discussed compilations (e.g. B11 and K13), and so far were not studied  in detail on an individual basis. Especially for poorly populated objects, however a detailed study is essential to confirm the reality as a cluster. In particular, for more distant open clusters, proper motion data do not significantly help to distinguish between cluster and field stars. Thus, the large surveys are probably contaminated by these kinds of  targets, influencing  statistical works on the Galactic cluster population, for example. 


\begin{figure*}
\centering
\includegraphics[width=170mm]{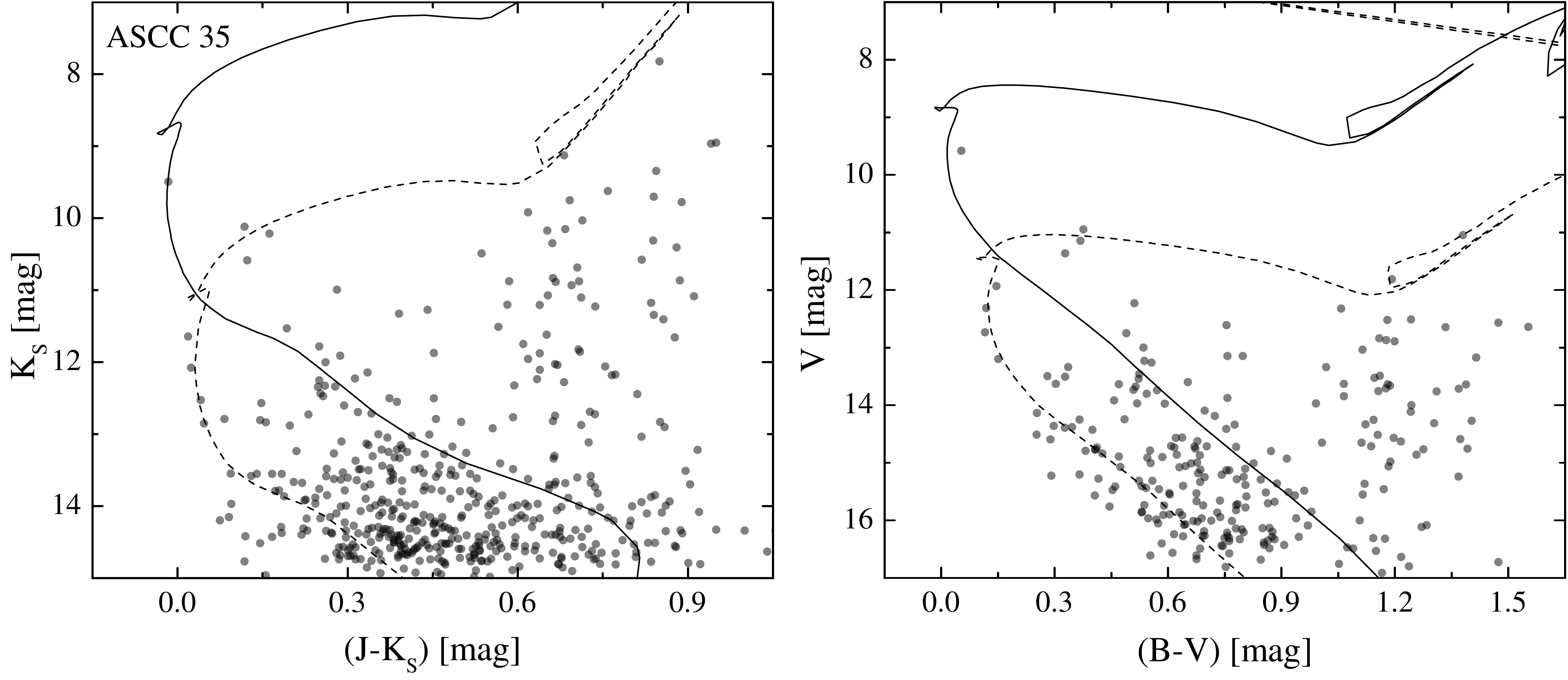}
\caption{The colour magnitude diagrams for ASCC~35 based on stars with kinematic membership probabilities larger than 60 percent. The solid line represents the isochrone with the parameters by K13 ($d = 787$\,pc, \ebv\ = 0.062\,mag, and \logt\ = 8.48), while the dashed line is an isochrone of same age but adopting a larger distance and a higher reddening ($d = 2200$\,pc and \ebv\ =0.17\,mag).} 
\label{fig:ascc35}
\end{figure*}


As an example of a poorly populated and little studied cluster candidate, we have chosen ASCC~35. This object was identified as an open cluster by K05 using ASCC--2.5 data. The target is included in the study by K13 as well, but with nearly identical parameters. K05 lists 800\,pc, 0.06\,mag, and \logt\ = 8.49\,dex for the distance, reddening, and age, respectively, while K13 derived 787\,pc, 0.062\,mag, and \logt\ = 8.48\,dex. The agreement of the parameters is implausible at first glance because of the differences of the used data sources (ASCC--2.5 and 2MASS) in respect of magnitude limit and wavelength. In the absence of any other data source in the visual besides the ASCC--2.5 catalogue, we used the AAVSO Photometric All-Sky Survey, Data Release 8 (APASS \footnote{http://www.aavso.org/apass}), which provides photometry in the bands $BVg'r'i'$ down to about $V$ $\approx$ 17\,mag. We performed a cross-match with the kinematic members listed by K13 and show the 2MASS and the visual CMD in Fig. \ref{fig:ascc35}. Only stars in the estimated cluster area ($r = 0\fdg235$, K13) are shown in both diagrams that fulfil the photometric and proper motion quality criteria by K13 and that have listed kinematic membership probabilities larger than 60 percent. There is hardly a cluster sequence visible, and the isochrone with the parameters listed by K13 (solid line) appears to be oriented to a single star.  

To demonstrate the difficulties in the isochrone fitting procedure, we also included in Fig. \ref{fig:ascc35}  an isochrone of the same age, but shifted towards a higher reddening and a larger distance (\ebv\ = 0.17\,mag and $d = 2200$\,pc). One could argue that these parameters provide a more reasonable fit because it covers more stars in the CMDs, however, by adopting the larger distance and the cluster diameter by K13, the absolute diameter amounts to about 18\,pc. Only young unbound groups such as associations have diameters of that size \citep{Jane88}. In particular for open cluster candidates such as ASCC~35, additional kinematic data are important to prove the reality as a cluster and to select the most probable member stars. Thus, the study of these targets will clearly benefit from the upcoming \textit{Gaia} data.

\section{Comparison of results for individual objects}
\label{sect:individual}
\begin{figure*}
\centering
\includegraphics[width=170mm]{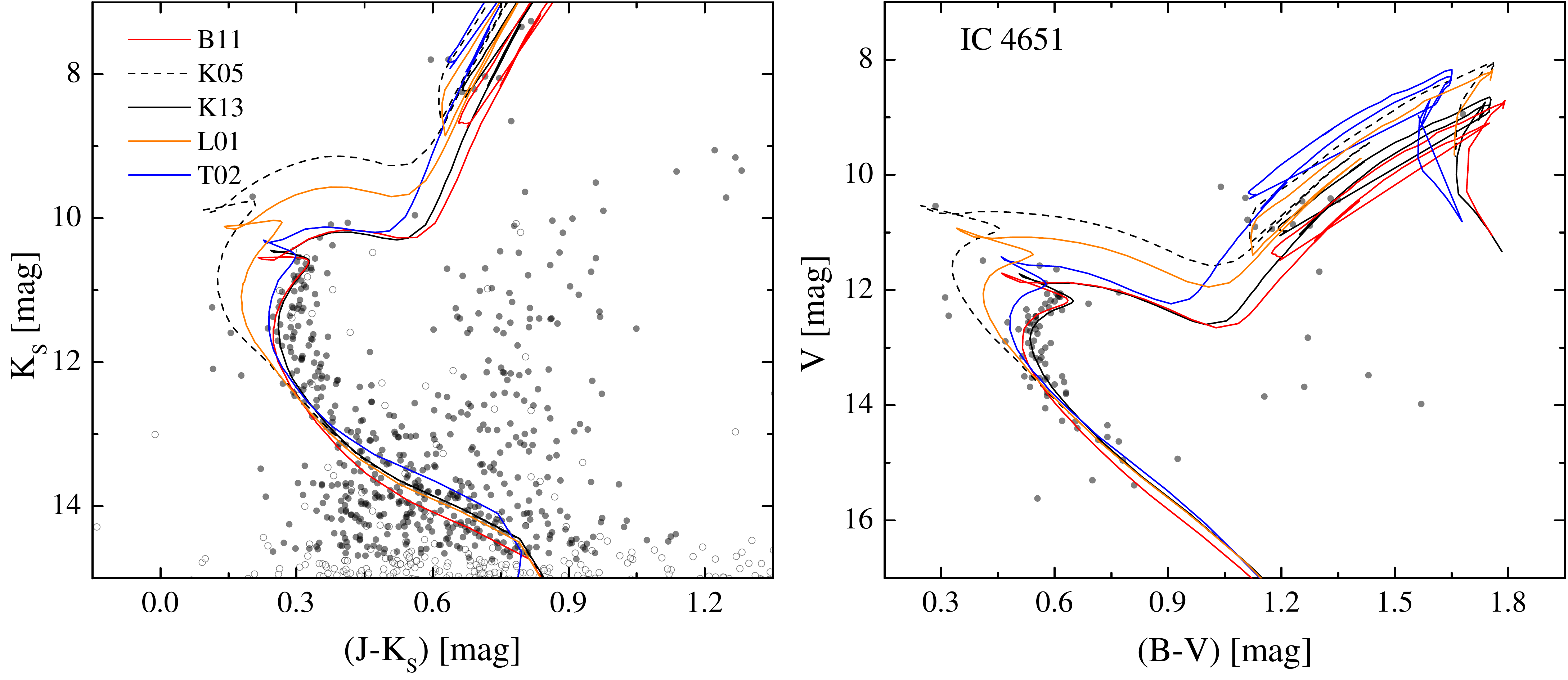}
\caption{The colour magnitude diagrams for IC~4651. The \bv\ data were taken from \citet{eggen71} and we used a radius of 5\arcmin\ around the cluster centre for both, the visual and NIR CMD. Filled symbols in the NIR CMD represent objects with a 2MASS photometric quality flag of A (S/N $>$ 10), while open symbols are stars that are covered with a lower photometric quality.} 
\label{fig:ic4651}
\end{figure*}

From the list of 54 open clusters that are included in at least five studies, we selected seven targets with sufficient \bv\  photometry. There is no object that is covered by all considered works, and only two clusters (NGC~2158 and NGC~7380) that are included in six different compilations. For the presentation of both the visual and NIR CMD, we used the inner parts of the open clusters. The radii were selected in a way to account for the often limited field of view of the visual (CCD) studies, to avoid overcrowding the CMDs and still cleary represent the cluster morphologies. The adopted values are listed in the captions of the Figs. \ref{fig:ic4651}, \ref{cmd1}, and \ref{cmd2}. Furthermore, we used a standard extinction law ($R=3.1$) to derive the apparent distance modulus. All NIR studies indicate the  reddening ratios used between 2MASS colours and \ebv, and for the studies in the visual we adopted \ejk/\ebv\ = 0.48 and $A_{K_\mathrm s}$=0.67\ejk\ to apply the distance and extinction to 2MASS isochrones. These ratios are those used by K13 (see Table \ref{tab:systcomp}). We used the isochrones by \citet{Mari08} for Z = 0.019, the metallicity that was adopted by most studies. The source for the \bv\ photometry was selected in respect of magnitude limit and coverage of the respective cluster. An overview of the cluster parameters is shown in Table \ref{tab:paracomp}. For completeness, we also list the cluster coordinates adopted or derived by the individual studies, and the cluster radii derived by K05 and K13. Some of the other studies also provide radii, but are defined in a different way and are therefore not directly comparable.

\textit{IC~4651}: This is the closest open cluster among the eight selected targets; it also one of the brightest open clusters. The visual and NIR CMD show both a well-defined MS, TO point, and red giants clump. Thus, most studies provide parameter sets that fit the CMDs very well. However, the TO identified by K05 is oriented to the bluest brighter stars, which results in a younger age compared to all other studies. Obviously, this is owing to the limiting magnitude of the ASCC--2.5 catalogue used by K05. The stars are probably blue stragglers, but only one is listed as bona fide by \citet{ahumada07}. The input values for reddening and distance adopted by K13 are those by K05, and have not changed after the fitting procedure. Actually, these are the results obtained by \citet{Lokt01}. The age obtained by L01 is somewhat too low as well, although some of the data sources available at that time provide  reasonable coverage of the cluster. A reddening of almost zero was derived by T02. However, \citet[][]{twarog00} used intrinsic colour relations for the $uvby\beta$ system and estimated \ebv\ $\approx$ 0.10\,mag, thus confirming the somewhat higher value listed by all of the other studies. Because of the lower reddening and the connection of all parameters in the isochrone fitting procedure, T02 also obtained  the smallest distance and the oldest cluster age among all studies.


\textit{NGC~2158}: 
The open cluster NGC~2158 is the faintest and also probably  the oldest object among the selected targets. All studies identified the distinctive red giants clump, but no investigation was able to reproduce all features of the cluster CMDs. Some results (L01 and T08) are again oriented to the bluest stars and do not match the TO at all, others adopt a  TO magnitude (B11 and G10) that is too bright, while the results by K13 and T02 do not fit the main sequence. The in-depth study in the visual by \citet{carraro02} estimated  \ebv\ = 0.55, thus the reddening is significantly higher compared to all listed studies in Table \ref{tab:paracomp}. This is related to the underabundant nature of the cluster, which was taken into account by \citet{carraro02}. \citet{heiter14} list an iron abundance [Fe/H] $\approx$ $-$0.3\,dex based on medium- and low-resolution spectroscopy. Nevertheless, the 2MASS data probably provide a  magnitude limit that is  somewhat too bright to estimate the parameters accurately.

\textit{NGC~2383}: The parameters listed in Table \ref{tab:paracomp} differ significantly especially for the age and the distance. The distances are between 1.7\,kpc and 3.5\,kpc, while the determined age spans values between  15\,Myr (L01) and almost 500\,Myr (K13). The studies by \citet{subramaniam99} and \citet{vazquez10} point towards an older age and  larger distance. Thus, the result by B11 appear to be the most reasonable, although the age is somewhat too large as can be noticed in the visual CMD. While the actual age is difficult to determine for the open cluster, depending on the consideration of possible red giants, the distances given by K05, K13, and L01 are definitely too small. As for several other objects, the distance and reddening values are the same in these three references. In particular, K13 should have noticed that the adopted starting value for the distance does not match the 2MASS data at all.

\textit{NGC~2489}: 
The radial velocities of a few giants \citep{piatti07} confirm that this cluster has a clump, and it is therefore of intermediate age. The automatic fitting by B11 (see Fig \ref{cmd1}) nicely
reproduces the CMD morphology, in line with detailed studies in the visual \citep{Neto13,piatti07,ramsay92}. All the other automatic fittings fail, suggesting the cluster is much younger.
Reddening goes from 0.37 to 0.74 mag, while distance goes from just 1.1 kpc (T02) all the way to 3.7 kpc (K05). The incorrect age derived from all the automatic fittings, other than B11, is most probably
related to the incapability of detecting the cluster clump, which in spite of being sparse and poor, is real. However, K13 provide tables for all open cluster with the photometric data and the derived membership probabilities for each star in the cluster area. We checked the results for NGC~2489 after restricting the star list to spatial members that meet the quality criteria for the photometric and proper motion data (see Sect. \ref{sect:compilations}). The red giant clump is still clearly visible, as in Fig. \ref{cmd1}, when using kinematic members with a probability larger than 80 percent. Thus, K13 has probably  relied more on the adopted input value for the age (taken from K05) instead of using an isochrone that fits the identified kinematic member stars. The parameters by L01 (and K05) do not provide a proper fit. We have not found any data set in the open cluster database WEBDA\footnote{http://webda.physics.muni.cz} that could explain the results by L01.

\textit{NGC~2627}:
This is another example of a well-studied cluster. Numerous works exist in the literature \citep{ahumada05,piatti03,ramsay92}, and all coincide that the cluster is of intermediate age.
However, automatic parameter assignments tend to yield younger ages. Only B11 provides a reasonable, but not completely correct, fit. K05, L01, and T02 completely miss the cluster. In general, the reddening
obtained is very low, still the distance varies from 1.5 to 2.7 kpc. One can argue that field star contamination  complicates the clear detection of the TO location. This might be true in NIR, but certainly not in the visual. The magnitude limit of the ASCC--2.5 catalogue is about 12\,mag in this field. Thus, the data used by K05 do not cover the cluster, except for a very small number of giants. A more careful check of the literature would have helped to identify that. Although the CCD data used by T02 \citep{ramsay92} are less numerous than that presented in Fig. \ref{cmd2}, they still cover the  features of the cluster down to about 18\,mag in $V$.
NGC 2627 is a clear example of automatic fitting failure.

\textit{NGC~6603}: 
This is an intermediate-age star cluster projected towards the Galactic bar/bulge, at only 12 degrees from the Galactic centre, and it stands as a precious example of a rich relatively old star cluster in the inner disk. In the literature there are a couple of good studies in the visual \citep{bica93,sagar98}.  Both the 2MASS and visual CMD in Fig. \ref{cmd2} clearly indicate we are facing an intermediate age cluster with a clump.
The MS termination point is blurred by field star contamination, which is expected to be severe in the Galactic direction to NGC 6603. Automatic fits in Fig. \ref{cmd2} produce reasonable fit, except for K05, which got 
clearly confused by bright interlopers from the Galactic field and ends up with untenable results. Excluding this particular case the reddening ranges from 0.6 to 0.8\,mag, distance from 1.6 to 3.5\,kpc, and age from 8.3 to 8.8
in logarithmic units. Looking at these fits more closely, it seems that in general reddening is overestimated in all cases, except for T02, who evidently estimated an age that is too old. An independent age indicator is the magnitude
difference between the TO and the clump, which, if respected, would have helped to get a more reasonable result for the age. K13 obtained a better solution of the age, but the reddening is too large. 
A by-eye examination confirms that none  of these fits correctly reproduce the cluster morphology.

\textit{Trumpler~14}: This object is one of the few young massive (starburst) star clusters in the Milky Way \citep[][]{beccari15,negueruela14}, and is located in the north-west extension of the Great Carina Nebula. Since it is a prominent cluster, it received much attention in the previous century.  Still, its fundamental parameters are far from being firmly settled. One clear complication, quite typical in young clusters, is that reddening is inhomogeneous across the cluster \citep{carraro04} . Additionally, the reddening law towards the Great Carina nebula does not seem to be normal \citep{turner12}. These two facts dramatically complicate  any study of the cluster aiming at inferring its fundamental parameters. Besides, Trumpler 14 contains
a huge population of contracting stars in the pre-MS phase \citep{ascenso07,sana10}. The age of stars in the pre-MS (contracting age) is not necessarily similar to the nuclear age, namely the age of stars that already are burning hydrogen on the main sequence \citep[see e.g.][]{baume03}.
Trumpler 14 is therefore a classical example of  a star cluster that needs to be considered very carefully before deriving its fundamental parameters, and the assistance of by-eye-inspection is crucial. 
\citet{beccari15} found a large age spread among pre-MS stars, confirming earlier findings by \citet{carraro04}. The age of the younger pre-MS component is, however, very close to the age of stars in the main sequence \citep[2--3\,Myr][]{vazquez96}.
If the distance to the cluster is inferred assuming a normal extinction law ($R=3.1$), and the same
reddening is adopted for every stars, this distance will be undoubtedly wrong. Star-to-star variations in reddening as large as 0.2 mag have  often been reported \citep[e.g.][]{carraro04}.
The plot in Fig. \ref{cmd2} illustrates very clearly that it is  difficult to decide where to pass an isochrone through the 2MASS data points. We refrain  from comment on this.
Optical data are better, since the main sequence is clearly visible. Now, it is readily clear that only the black dashed isochrone (K05) is producing a reasonable fit, although reddening is clearly too large.
With this reddening, an age smaller than \logt\ = 6.67 would help. 
One of the typical mistakes when trying to fit isochrone onto young star cluster sequences is that
much attention is paid to the main sequence below the TO point of the pre-MS (V $\approx$ 15.0\,mag). This is wrong because that region of the CMD is contaminated by interlopers, and most star members are located red-ward, in the pre-MS. If attention is paid exclusively to the main sequence, more solid estimates of distance and reddening are possible. Table \ref{tab:paracomp} show reddening estimates from 0.45 to 0.87 and distance estimates from 2.2 to 2.7 kpc. Age ranges from 1 to 12 Myr.

We note that some of the selected objects show peculiarities. K13 state in their catalogue that NGC~2158, NGC~2383, and Trumpler~14 overlap with other open clusters, however, only the outskirts of the clusters are influenced by this fact. The central parts, which are used, e.g. by K13 for their analysis, are well separated or only marginally affected.


\begin{table*}
\caption{The parameters for the selected open clusters.} 
\label{tab:paracomp} 
\centering 
\begin{tabular}{l l c c c c l} 
\hline\hline 
Cluster & $\log t$ & $d$ & $E(B-V)$ & $\alpha$ / $\delta$ (2000) & radius $r_1$/$r_2$\tablefootmark{c}\\ 
        &          & [pc] & [mag] & hh:mm:ss / dd:mm:ss & [\arcmin] \\
\hline 
IC~4651 (5\arcmin) & 9.15 & 1004 & 0.15 & 17:24:47/$-$49:56:18 &   & B11 \tablefootmark{a}\\
                & 8.92 & 888 & 0.12 & 17:24:50/$-$49:55:48 & 5.4/10.8 & K05 \tablefootmark{b}\\
                & 9.25 & 888 & 0.121 & 17:24:54/$-$49:56:06 & 9.6/14.4 & K13 \tablefootmark{a}\\
                & 9.057 & 888 & 0.116 & & & L01 \tablefootmark{b}\\                     
                & 9.34 & 771 & 0.02 & 17:24:42/$-$49:55:00 &  & T02 \tablefootmark{b}\\
NGC~2158 (1\farcm5)& 9.10 & 4058 & 0.44 & 06:07:26/+24:05:46 &   & B11 \tablefootmark{a}\\
                & 9.30 & 3300 & 0.34 & 06:07:28/+24:05:53 & & G10 \tablefootmark{a}\\
                & 9.33 & 4770 & 0.333 & 06:07:26/+24:05:31 & 8.4/15.0 & K13 \tablefootmark{a}\\
                & 9.023& 5071 & 0.36 & & & L01 \tablefootmark{b}\\              
                & 9.20 & 5012 & 0.40 & 06:07:24/+24:05:00 &   & T02 \tablefootmark{b}\\
                & 9.00 & 4980 & 0.35 & 06:07:25/+24:05:48 & & T08 \tablefootmark{a}\\
NGC~2383 (2\farcm5)& 8.60 & 3494 & 0.31 & 07:24:41/$-$20:56:43 &   & B11 \tablefootmark{a}\\
                & 8.64 & 1655 & 0.21 & 07:24:38/$-$20:57:00 & 3.6/6.0 & K05 \tablefootmark{b}\\
                & 8.69 & 1655 & 0.21 & 07:24:41/$-$20:56:24 & 4.2/7.5 & K13 \tablefootmark{a}\\
                & 7.167& 1655 & 0.213& & & L01 \tablefootmark{b}\\              
                & 7.60 & 3048 & 0.28 & 07:24:42/$-$20:55:00 &   & T02 \tablefootmark{b}\\
NGC~2489 (5\arcmin)& 8.45 & 1846 & 0.44 & 07:56:17/$-$30:03:26 &   & B11 \tablefootmark{a}\\
                & 7.25 & 3700 & 0.37 & 07:56:17/$-$30:03:36 & 3.6/6.6 & K05 \tablefootmark{b}\\
                & 7.315 & 2255 & 0.729 & 07:56:18/$-$30:03:58 & 4.2/7.5 & K13 \tablefootmark{a}\\
                & 7.264& 3957 & 0.374& & & L01 \tablefootmark{b}\\
                & 8.25 & 1148 & 0.40 & 07:56:12/$-$30:03:00 &   & T02 \tablefootmark{b}\\
NGC~2627 (4\arcmin)& 9.15 & 1871 & 0.06 & 08:37:13/$-$29:57:50 &   & B11 \tablefootmark{a}\\
                & 8.71 & 2034 & 0.09 & 08:37:14/$-$29:57:00 & 4.8/7.8 &  K05 \tablefootmark{b}\\
                & 9.225 & 2712 & 0.104 & 08:37:19/$-$29:56:42 & 5.4/9.9 & K13 \tablefootmark{a}\\
                & 8.566& 2034 & 0.086& & & L01 \tablefootmark{b}\\
                & 8.60 & 1515 & 0.05 & 08:37:18/$-$29:56:00 &   & T02 \tablefootmark{b}\\
NGC~6603 (2\arcmin)& 8.75 & 1900 & 0.62 & 18:18:28/$-$18:24:34 &   & B11 \tablefootmark{a}\\
                & 7.80 & 2880 & 0.50 & 18:18:24/$-$18:25:48 & 3.6/7.2 & K05 \tablefootmark{b}\\
                & 8.40 & 2325 & 0.833 & 18:18:25/$-$18:25:48 & 4.2/8.4 & K13 \tablefootmark{a}\\ 
                & 8.80 & 1570 & 0.56 & 18:18:30/$-$18:24:00 &   & T02 \tablefootmark{b}\\
                & 8.30 & 3495 & 0.77 & 18:18:26/$-$18:24:24 & & T08 \tablefootmark{a}\\
Trumpler~14 (2\farcm5) & 6.85 & 2249 & 0.87 & 10:43:46/$-$59:33:45 &   & B11 \tablefootmark{a}\\
                        & 6.67 & 2753 & 0.45 & 10:43:55/$-$59:33:00 & 4.2/6.0 & K05 \tablefootmark{b}\\
                        & 6.00 & 2248 & 0.75 & 10:43:55/$-$59:33:18 & 4.2/7.8 & K13 \tablefootmark{a}\\
                        & 6.828& 2733 & 0.516& & & L01 \tablefootmark{b}\\
                        & 7.10 & 2427 & 0.50 & 10:43:56/$-$59:33:00 &   & T02 \tablefootmark{b}\\                 
\hline 
\end{tabular}
\tablefoot{The radii used for the CMDs in Figs. \ref{fig:ic4651}, \ref{cmd1}, and \ref{cmd2} are given in parentheses next to the object designations. 
\tablefoottext{a}{The parameters were derived using NIR 2MASS data.}
\tablefoottext{b}{The parameters were derived using data in the visual.}
\tablefoottext{c}{$r_1$: The radius where the decrease
of stellar surface density stops abruptly; $r_2$: The radius where the surface density of stars becomes equal to the average density of the surrounding field.}

}
\end{table*}
  

\subsection{Differences in the isochrone fitting procedures}
\label{sect:diffiso}
For all the objects discussed in Sect. \ref{sect:individual}, the open cluster parameters by L01 do not provide a proper fit to the CMDs. We cannot definitely conclude if this is due to the applied method or owing to the available data material. One reason could be that most of the selected objects are too distant and, therefore, too faint to be sufficiently covered by photometry at that time. However, this is clearly not the case for IC~4651 (Fig. \ref{fig:ic4651}), for which even the photoelectric \ubv\ data by \citet{eggen71} provide a reasonable coverage of the cluster. L01 used the oldest isochrone set among all the studies, but there are no significant differences between the \ubv\ isochrones by \citet{Bert94} and those adopted here (see discussion in Sect. \ref{sect:compilations}). Thus, the reason for the discrepancies might be due to the merging of individual results and the weighting scheme. 

The study by K05 adopted the distance and reddening estimates by L01 and other references for many open clusters and concentrated on the estimation of the age. Thus, their results are affected similar to those of L01. Here, the magnitude limit of the ASCC--2.5 catalogue clearly prevents better results. For many of the clusters, the isochrones fit the brightest and bluest stars in the cluster area. These are, however, either blue stragglers or even non-member stars (see e.g. IC~4651 or NGC~2627). Hence, K05 probably underestimates the age for several objects.

T02 used available \ubv\ CCD photometry for their investigation. The most recent photometric studies adopted by T02 date back to 1999, and many data originate from the very beginning of the CCD era. Thus, besides the considerably limited field-of-view of the first CCDs, the photometric accuracy  probably varies  strongly from one to another cluster. Our references for the photometric data differ compared to T02 for all objects, but NGC~2383.  However, even when using the photometry adopted by T02, we cannot conclude which fitting strategy was employed. Some results are oriented to the brightest stars or envelope the bluest part of the cluster sequence (NGC~2383 or IC~4651), while other isochrone fits are somehow placed  in the middle of the cluster sequences (e.g NGC~2489 or Trumpler~14). 

As mentioned in Sect. \ref{sect:compilations}, in their series
of papers T08 varied  the criteria to select the 2MASS data as well as the interstellar extinction ratio. The target selection is biased towards little studied objects, thus there is only a small overlap with most of the other studies. Besides the two clusters discussed in Sect.
\ref{sect:individual}, we also  examined  some additional clusters from the compilation by T08. Based on the two clusters NGC~2158 and NGC~6603 shown in Figs. \ref{cmd1} and \ref{cmd2}, it already becomes clear that the isochrone envelopes the bluest and brightest parts of the 2MASS CMD. However, this procedure seems to fail when inspecting the visual CMD of NGC~2158, where the adopted TO point is significantly off from the cluster. For this cluster, one is again faced with a mixture of field star contamination and a large number of blue stragglers \citep[40 candidates,][]{ahumada07}. The derived distance and reddening is comparable to most other studies, but the fitting procedure results in an  age that is too young. There is a noticeable trend towards the distance compared to K13 (see Sect. \ref{sect:global}). We therefore inspected the 2MASS CMD of one of the most distant objects in common (Berkeley~1). K13 estimated a distance of 10.3\,kpc, while T08 lists 2.4\,kpc (B11 quote 3.4\,kpc and T02 even 1\,kpc). The isochrone with the parameters by T08 again form an envelope around the bluest and brightest stars in the area, whereas K13 fitted an isochrone through the centre of the distribution of the stars in the CMD. These different approaches could explain the discrepancy in distance between the two references for several clusters (or candidates). We note that \citet{Phel94} question the reality of Berkeley~1 based on \ubv\ photometry.

The 2MASS results by B11, in general, show the best agreement if we also consider the visual region. For example, it is the only work that identified the red giant clump in the CMD of NGC~2489. Also, for NGC~2627, the parameters fit best in the visual. Their isochrone fitting approach is comparable to that by T08 (oriented to the bluest, brightest stars), but apparently is not as strict as that done by T08, resulting in a better agreement in the visual CMD. However, there are also some results that do not fit correctly (e.g. NGC~2158 or Trumpler~14). Some objects were investigated by B11, but  were excluded in the larger survey by K13. For example, K13 note for Reiland~1 that it is a compact asterism. Furthermore, B11 includes IC~2156 and IC~2157, but, according to K13, IC 2156 is either a clump in the corona of IC~2157 or an asterism. 

The study by G10 presents a sample that is of comparable size to that by T02, but focussed on cluster candidates they detected in their study. Thus, there is only a small overlap with other studies beside K13. Among the selected open clusters, only NGC~2158 is covered by G10. Based on this it is obvious that the fitting approach is comparable to B11. This is confirmed if we inspect some of the diagrams presented on their catalogue webpage \footnote{http://ocl.sai.msu.ru}.

The 2MASS study by K13 provides the largest sample of open cluster parameters, and all seven selected clusters are included in this survey. They mention that the additional use of kinematic data probably makes their study  more reliable compared to other purely photometric approaches. This is certainly true, if kinematic members are actually considered in the fitting solution. The cluster NGC~2489 is a nice counter-example. It seems that they place the isochrone, in general, through the centre of the distribution of the stars in the CMD. Good examples for this approach are NGC~2158, NGC~2627, and NGC~6603, but also Berkeley~1, which we discussed above. However, these results do not fit in the visual. On the other hand, for example, in the case of NGC~2383 K13 do not follow this approach. Here, it appears that they relied too much on the adopted input parameters for  distance and  reddening. For objects in common with K05, they used their previous catalogue for the input values. We noticed that these parameters remain in principle unchanged for about 15 percent of the targets in common, even for targets more distant than 1.5\,kpc. As criteria we used 0.05\,mag for the true distance modulus and 0.01\,mag for the reddening, values that can be easily achieved by rounding and the transformation between the visual and NIR. The input values that originate from K05 are in most cases the results by L01 (see Sect. \ref{sect:compilations} and discussion above). Thus, the comparison with the results by L01 cannot be considered completely independent.

\section{Corrections of the open cluster parameters}
\label{sect:recal}

In Sect. \ref{sect:global} we presented some basic deviations of the open cluster parameters compared to the data set by K13. Figure \ref{fig:statistics} shows that there are numerous objects with strong deviating results, which could influence a detailed calculation of correction terms between these samples. Furthermore, any homogeneous compilation can show some systematic trends, depending on the applied method and approach. 
Therefore, independent results are essential to further validate the cluster parameter surveys. We consider the results by individual studies to be  an independent reference sample, thus works that deal with a detailed analysis of a single open cluster or a small number of objects. This sample certainly incorporates different methods and approaches. Thus, dependencies in this  comparison must originate from the respective method of the surveys.

\citet{Pau06} studied the accuracy of open cluster parameters. Besides the use of some surveys also considered  in the present work (e.g. K05), they also compiled a comprehensive list of open cluster parameters derived by individual studies. All references are available in WEBDA\footnote{http://webda.physics.muni.cz/recent\_data.html}. From this list, we adopted results published since 1995 (351 parameter sets for 263 clusters). This should, on the one hand, somehow assure the availability of previous studies, and thus a broader range of data types and information for the cluster that can be additionally used to derive accurate results, and, on the other hand, guarantee the availability of modern evolutionary models, like those discussed in Sect. \ref{sect:compilations}. We updated this parameter compilation with more recent studies, resulting in an incomplete final list with more than 800 cluster parameter sets for about 450 open clusters, taken from almost 300 references. Note that we excluded works if an author of one of the parameter compilations was included. All the additional references and parameters will be made available in WEBDA as well. 

We derived mean values for the parameters if more than one result was available. The number of estimates for a single parameter may vary because some compiled references have not derived the complete parameter set \citep[e.g.][who presented mean parallaxes]{leeuwen09}. For 179 open clusters there are at least two estimates for each parameter available. In the following, we  use the sample with mean cluster parameters as calibrators for the results of the surveys. This data set probably provides an improved accuracy with an estimate of the errors,. Most importantly, however, the data set provides a sample of true open clusters that are confirmed by at least two individual studies. We use the other objects of the reference sample with single parameter determinations just for additional guidance purposes.  

\begin{figure*}
\centering
\includegraphics[width=170mm]{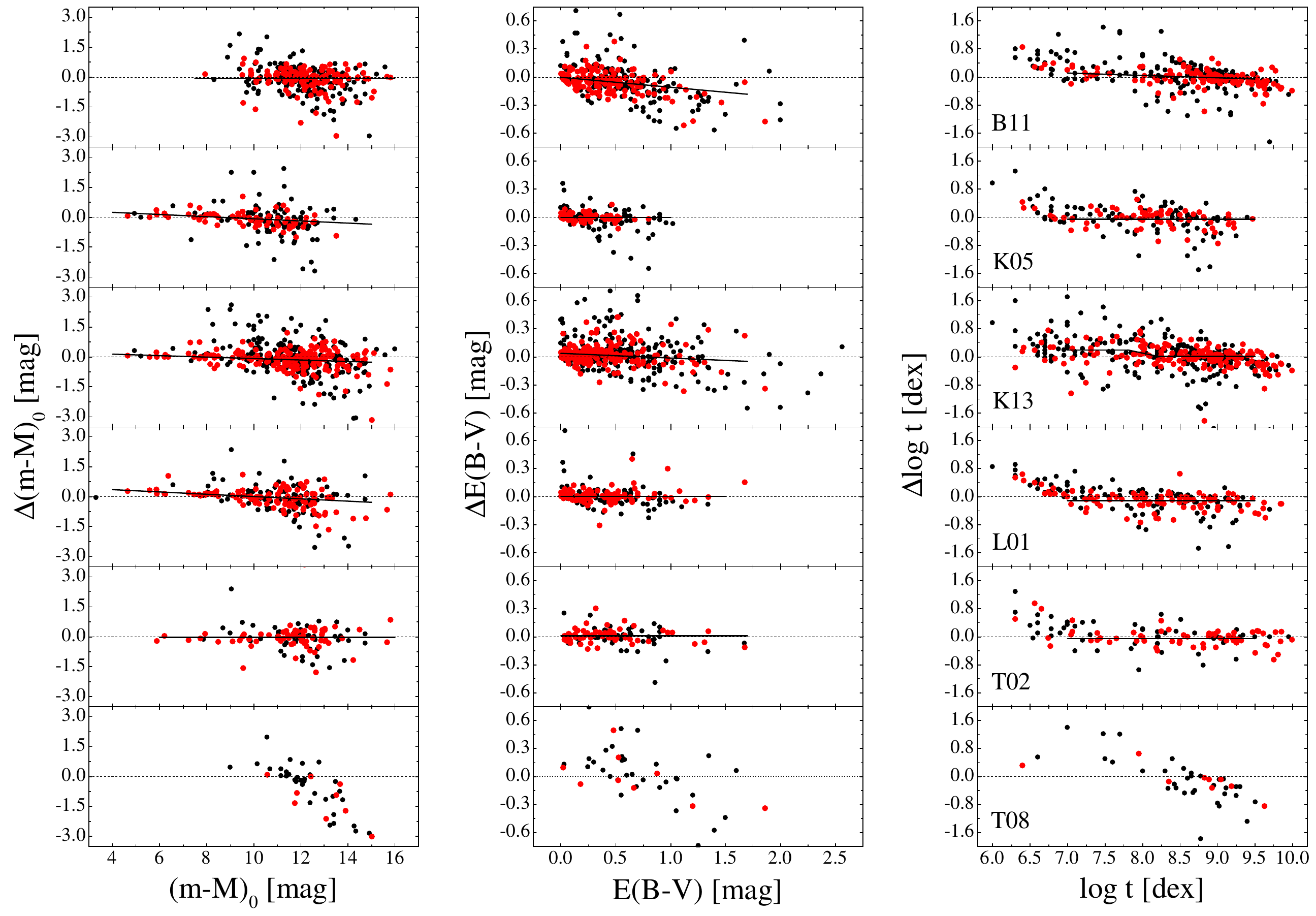}
\caption{Comparison of the parameters derived by the surveys with our reference sample. The differences (survey $-$ reference) are shown as a function of the reference values. Red symbols indicate the differences in the calibration sample, while black symbols are the differences in the remaining reference (single value) sample. } 
\label{fig:calibration}
\end{figure*}

The comparison between the survey results and the calibrators is shown in Fig. \ref{fig:calibration}. Compared to Fig. \ref{fig:statistics},  even visually a much lower scatter is noticeable for most surveys. We exclude G10 for now because of the very small overlap with the reference sample (three calibrators, 12 objects in total). Although the study by T08 shows a small overlap as well (about 40 clusters in total), one can conclude that these results appear much more erroneous than the others. The clear trend with the distance, already noticed in Fig. \ref{fig:statistics}, is confirmed to be due to their fitting procedure (see discussion in Sect. \ref{sect:diffiso}). A correction of these results is impracticable because the dependency is owed to the roughly equal distances estimated by T08 for most objects. Comparable effects can also be concluded  for the other two parameters. The strong deviating results are not restricted to a single paper of their series. We therefore do not consider this data set in the subsequent analysis. 


The comparison of the age indicates that most surveys deviate at the young and old tail of the distribution. All of the studies other than K13 use isochrones starting with \logt\ = 6.6, resulting in larger discrepancies towards the very young end. However, the deviation already seems to start  at a somewhat older age (\logt\ $\lesssim$ 7.0), which is most noticeable in the data by L01 or B11. The parameter determination for very young open clusters is in general affected by some difficulties, such as strong differential reddening. Furthermore, the studies by B11, K13, and L01 seem to underestimate the age  of open clusters older than \logt\ $\sim$ 9.5 by about 0.3\,dex, although the used isochrones are not limited as was the case at the young border. One could argue that the discrepancy is a result of the use of solar metallicity isochrones by the surveys. However, a comparison with mean spectroscopic iron abundances \citep{heiter14} shows that the underestimation of the age can be found for underabundant clusters (e.g. Berkeley~20 or Trumpler~5),  for objects that show roughly solar metallicity (Berkeley~17, Collinder~261), or for overabundant clusters (NGC~6791). An inspection of some NIR CMDs in the online catalogue by K13 shows that the magnitude limit of the 2MASS data might be a problem for very old and more distant ($\gtrsim$\,2\,kpc) open clusters because only the red giants branch and a small portion of the TO (if any) are covered.

Thus, we adopt the age range 7.0 $\leq$ \logt\ $\leq$ 9.5 to obtain somewhat unbiased samples for which  the results can be directly compared. The studies by L01 and T02 used higher values for $R_V$ (see Table \ref{tab:systcomp}), and we corrected the distance modulus  using $R_V = 3.1$ and their results for the reddening. However, to reduce the influence of this correction, we use only open clusters for the distance calibration that show a reddening lower than 0.7\,mag (T02) or 0.5\,mag (L01). These reddening limits correspond to maximal differences in the true distance modulus of about 0.1\,mag, owing to the change of $R_V$. Finally, we derived corrections for the parameters in the form $\Delta$ = a + b\,X, with X as the parameter value of the calibration sample and $\Delta$ as the difference between the survey results and the calibrators. For some parameters a constant offset appears to be the most reasonable solution. The results by T02, for example, need only marginal, constant corrections for all three parameters. Table \ref{tab:recal} lists all corresponding values and the number of objects used for the calculation. 


\begin{table*}
\caption{Correction terms for the cluster parameters}
\label{tab:recal}
\centering
\begin{tabular}{cllll|lll|c}
\hline
\hline 
  & \multicolumn{2}{c}{$\Delta (m-M)_0$ [mag]} & \multicolumn{2}{c|}{$\Delta E(B-V)$ [mag]} & \multicolumn{3}{c|}{$\Delta$log\,\textit{t} [dex]} & N \tablefootmark{a} \\
  & \multicolumn{1}{c}{a} & \multicolumn{1}{c}{b} & \multicolumn{1}{c}{a} & \multicolumn{1}{c|}{b}  & & \multicolumn{1}{c}{a} & \multicolumn{1}{c|}{b} &  \\
\hline
B11     & $-$0.058(330) &  &  $-$0.001(16) & $-$0.108(29)  & 7.0 $\leq$ X $\leq$ 9.5 & +0.608(226) & $-$0.070(26)   &       101/104/103\\
G10     & +1.797(598) & $-$0.166(47)   & +0.027(38)   & $-$0.144(42) &   8.3 $\leq$ X $\leq$ 9.5 & $-$0.007(220) & & 98/98/98   \\
K05     & +0.447(155) & $-$0.053(16)    &  $-$0.003(32) &  &  7.0 $\leq$ X $\leq$ 9.5 & $-$0.049(194) &    &   63/62/56    \\
K13     & +0.278(158) & $-$0.036(14)    & +0.039(11) & $-$0.052(23)   & 7.0 $\leq$ X $\leq$ 8.0 & +0.195(286) &   &  148/140/30    \\
        &  &   &  &  &  8.0 $<$ X $\leq$ 9.5 & +0.014(182) &    & 111      \\
        &  &   &  &  &  7.8 $\leq$ X $\leq$ 8.2 \tablefootmark{b} & +3.725 & $-$0.453   &       \\
L01     & +0.591(194) & $-$0.058(19)    & $+$0.002(46) &  &  7.0 $\leq$ X $\leq$ 9.5  & $-$0.113(158) &   & 67/85/77      \\
T02     & $-$0.029(283) &   &  +0.010(62) &  &  7.0 $\leq$ X $\leq$ 9.5  &  $-$0.051(145) &   &  41/48/44 \\
\hline
\end{tabular}
\tablefoot{The coefficients of a regression $\Delta$(survey$-$cal) = a + b\,X, with X as the parameter value of the calibration (cal) sample, or constant offsets. The errors of the last significant digits are given in parentheses.
\tablefoottext{a}{The number of objects used for the calibration of the individual parameters.} \tablefoottext{b}{We suggest  applying this correction to allow a smooth transition between the plateaus.}}
\end{table*}


In general, most parameters are reasonably scaled. However, an interesting feature was found in the comparison with K13. While the age of clusters older than about 100\,Myr are in agreement to the reference sample, K13 overestimates the age of younger objects by almost 0.2\,dex. We are not able to identify the reason for these differences, but it might be related to the  method used. While K13 determined the age of younger clusters with the isochrone-fitting technique, for older objects the mean age of the TO stars was derived. As given in Table \ref{tab:recal}, we suggest  applying an additional correction around 100\,Myr to allow a smooth transition between the two age plateaus.

Figure \ref{fig:calibration} shows a clear trend in the reddening derived by B11. This might be the result of the isochrone-fitting approach, but also caused by a faulty transformation to \ebv.
As listed in Table \ref{tab:systcomp}, the 2MASS studies by B11 and K13 used different ratios to convert \ejk\ to \ebv. The lower ratio adopted by K13 shows better agreement with the reference sample, but still appears to be  too large. Both data sets suggest a comparable extinction ratio \ejk/\ebv: 0.455$\pm$0.011 (K13) and 0.464$\pm$0.015 (B11). The latter is almost identical to the ratio derived by B11 in their comparison with the open cluster catalogue by \citet{Dias02}. Both values agree well with the result of 0.466 by \citet{yuan13}, who used a synthetic stellar spectral model and the extinction law of \citet{Card89}, however, they derived empirical coefficients as well,  which indicate a much lower ratio of 0.414 $\pm$ 0.010.


\begin{figure*}
\centering
\includegraphics[width=170mm]{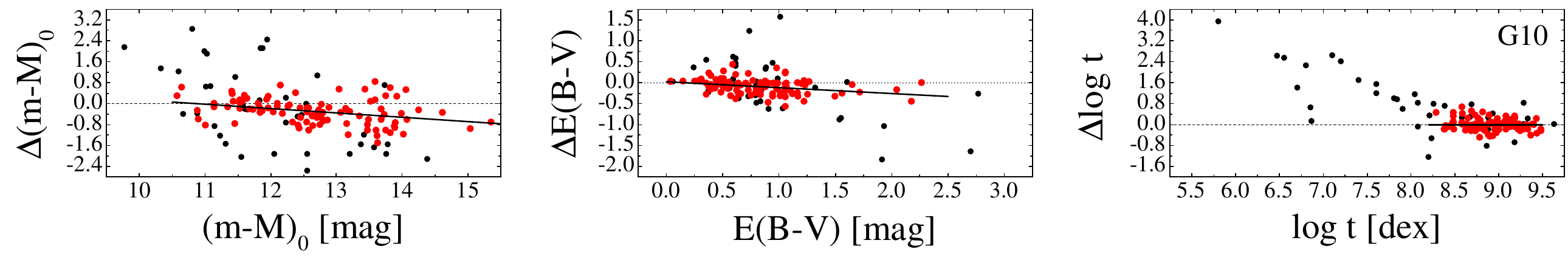}
\caption{Comparison of the parameters derived by G10 to the corrected results by K13. The differences (G10 $-$ K13) are shown as a function of the results by K13. Red symbols indicate the selected calibration sample.} 
\label{fig:g10cal}
\end{figure*}


As mentioned earlier, the data set by G10 marginally overlaps with the reference sample. Thus, we aim to verify it with the corrected results by K13. However, this comparison probably also includes  doubtful or poorly populated objects in which unique fitting solutions are hindered. For example, Fig. \ref{fig:statistics} shows that several young objects, according to K13, are old objects in the compilation by G10. However, for objects older than \logt\ $\sim$ 8.3 the results seem to agree. We therefore used only the older objects to derive the corrections listed in Table \ref{tab:recal} and shown in Fig. \ref{fig:g10cal}. Objects that deviate significantly in a single parameter were excluded completely. The results are certainly affected by this selection and have to be used with caution. A more detailed analysis has to be performed as soon more objects of the sample by G10 are independently confirmed and also analysed by individual studies. The results by G10 seem to underestimate the distance for more distant objects (about 600\,pc at a distance of 4\,kpc), but there is also a tendency with reddening comparable with B11. The reddening by G10 was derived using \ejh\ and the ratio \ejh/\ebv = 0.33. Thus, the difference could be owed to an incorrect ratio as well. The data would suggest 0.282 $\pm$ 0.014, which is again in agreement with the value 0.279, derived by \citet{yuan13} based on the Cardelli reddening law. However, the direct comparison of \ejh\ with the results tabulated by K13 shows a tendency as well: $\Delta$\ejh\ = $-$0.003 $-$0.070\ejh. Thus, the found differences are more related to the fitting approach.  


\begin{table}
\caption{Errors and statistical properties of the corrected parameters.} 
\label{tab:statrecal} 
\centering 
\begin{tabular}{l c c c c c } 
\hline\hline 
   & $\sigma(m-M)_0$ & $\sigma E(B-V)$ & $\sigma \log t$ &  N & AF$_{\rm Cal}$ \\ 
   & [mag] &  [mag] &  [dex] &   &  \\ 
\hline 
B11 &   0.44    & 0.14 & 0.22   &  105 &  56  \\        
    &  \textit{0.26}    & \textit{0.07} & \textit{0.14}    &   &    \\
G10 &   0.55    & 0.23 & 0.26   &  98 &  50   \\        
    
K05 &   0.33    & 0.05 &0.25    &  60 &  48  \\ 
    &   \textit{0.18}    & \textit{0.03} & \textit{0.14}    &   &   \\ 
K13 &    0.49    & 0.12 & 0.31  & 144  &  56  \\
    &  \textit{0.25}    & \textit{0.06} & \textit{0.14}    &   &    \\
L01 &   0.46    & 0.07 &0.22    &  82 &  55   \\
    &   \textit{0.18}    & \textit{0.04} &\textit{0.12}    &   &   \\
T02 &   0.40    & 0.08 &0.19    &  46 & 50   \\
    &   \textit{0.21}    & \textit{0.05} &\textit{0.12}    &      &  \\

\hline 
\end{tabular}
\tablefoot{
The mean errors of the corrected parameters in the adopted age range and the number of objects (N) used for the calculation. The second line in italic style gives the mean error of the respective calibration sample itself (note that G10 was compared to K13). The agreement factor was calculated for objects in common with the calibration sample (AF$_{\rm Cal}$) for the age range 7.0 $\leq$ \logt\ $\leq$ 9.5.
}
\end{table}


Table \ref{tab:statrecal} lists the mean errors of all corrected parameters, which are derived using the calibration samples in the adopted age range (including previously rejected outliers) as the basis. We assume that the errors of the surveys and the calibration sample are added in quadrature. Therefore, we also derived  the mean error of the calibration sample using the same objects (given in italic style). After the correction, all  of the surveys show a comparable accuracy and a mean intrinsic error of about 0.2\,dex for the age, 0.08\,mag for the reddening, and 0.35\,mag for the distance modulus. However, the visual studies seem to have a  performance that is twice as good for  reddening than the 2MASS surveys. This indicates that this parameter is probably more difficult to estimate in the NIR. The study by K13 shows the largest error for the age ($\sim$ 0.28\,dex) and together with L01 also the largest error for the distance. In this comparison, we do not consider G10 because of the two-step correction. Table \ref{tab:statrecal} shows that the age, for example, has a lower error than the results by G13, although both errors should add in quadrature. 

In agreement with the conclusion in Sect. \ref{sect:diffiso}, the results by B11 are among the most accurate, in particular, if directly compared with the other large survey by K13. Surprisingly, the work by T02 shows slightly smaller errors than B11, although the comparison of the individual objects in Sect. \ref{sect:individual} would not indicate that. The selected clusters are therefore probably not a perfect gauge for the average performance, but they still show up the difficulties to obtain proper cluster parameters.

We again use the agreement factor AF, already introduced in Sect. \ref{sect:global}, as a measure of the overall agreement of the results. With the individual errors at hand, we first derive the AF of the surveys using the calibration sample. This provides us with a reference value, the percentage that can be expected for a sample that consists of true open clusters. This AF$_{\rm Cal}$ is almost identical for all surveys and amounts to about 50\,\% (see Table \ref{tab:statrecal}). The adopted errors are about half the values used in Sect. \ref{sect:global} and Table \ref{tab:distributions}. Finally, we repeat the calculation of the agreement factor by comparing the corrected survey results with K13 (AF$_{\rm K13}$). As a limit, we use the individual parameter errors, which were added in quadrature after removing the error contribution of the calibration samples. On average, the limits are about 0.13\,mag for \ebv, 0.55\,mag for the distance modulus and 0.33\,dex for the age. Furthermore, we use the corrected age by K13 to restrict the samples to the adopted age range 7.0 $\leq$ \logt\ $\leq$ 9.5. One might expect comparable values for AF$_{\rm Cal}$ and AF$_{\rm K13}$, unless the samples include a larger percentage of problematic cases. These could be poorly populated objects with an ambiguous CMD (see e.g. ASCC~35 in Fig. \ref{fig:ascc35}), clusters with a high degree of field star contamination, or even doubtful open clusters. 

While there is a rough agreement between both AF values for the studies by L01 and T02, the others show larger deviations. The data by K05 result in an unexpected increase of the agreement factor by 14\,\%, which might be the result of the close connection between the surveys by K05 and K13. We already discussed in Sect. \ref{sect:diffiso} that for about 15\,\% of the objects the results have not changed between the studies (at least distance and reddening). On the other hand, the agreement between the large and independent surveys by B11 and K13 is 23\,\% lower than AF$_{\rm Cal}$. The difference would be even larger when also including  objects younger than 10\,Myr because K13 suggest a very young age for numerous objects, while B11 estimate an old nature (see Fig. \ref{fig:statistics}). Thus, the sample by B11 probably includes at least 20\,\% of problematic objects. It is very likely that this value, or even a larger one, also holds for K13, who presented parameters for almost four times more objects, the majority of them not studied in detail so far. A similar conclusion could be drawn for the sample by G10 (AF$_{\rm K13} = 38$). However, the samples used for these calculations are biased because the results by K13 were used to derive both AF values.  

\section{Summary and conclusions}
\label{sect:conclusion}

We presented a comparison of the open cluster fundamental parameters: distance, age, and reddening, compiled from seven different studies. These parameters represent results based on different photometric data and/or approaches to analyse the characteristics of open clusters. 

First of all, we performed a comparison in respect to the largest study by K13 to derive some basic statistical properties. The mean standard deviation for the distance, age, and reddening amounts to 1.2\,kpc (or 0.8\,mag for the true distance modulus), 0.5\,dex, and 0.27\,mag, respectively. For some objects, the results differ even by 8\,kpc for the distance or 3\,dex for \logt. At first sight, the large errors are clearly disappointing because open clusters are routinely considered as {\it \textup{bona fide}} objects 
for which it is relatively easy to derive accurate parameters. Some studies show clear trends in the distance or  age (see Fig. \ref{fig:statistics}).  In particular, the errors for the reddening, in general the best-known parameter for open clusters \citep{Pau06}, appear to be quite large. This error,  translated into  total absorption $A_{\mathrm V}$,  already amounts to about 1\,mag. Thus, one has to carefully check a cluster parameter set, if it is used to derive photometric temperatures, luminosities for individual cluster stars, or a luminosity function of an open cluster, for example.

Furthermore, we used a selection of open clusters to identify the isochrone-fitting strategies. We found that no study was able to reproduce the cluster morphologies of all the selected clusters (IC~4651, NGC~2158, NGC~2383, NGC 2489, NGC 2627, NGC~6603, and  Trumpler~14) in a correct and consistent manner. However, we identified the study by B11 as the most reliable, at least for this small sample. To identify difficulties in the respective methods, a much larger number of open clusters must be analysed in a similar way. This sample must cover the complete age and distance range and a varying field star contamination, with several already well-studied objects in each group. Despite the difficulties in finding this kind of  sample, with a growing number less time can be spent to interpret an individual object and the respective results. The small amount of time that can be spent on a single target is probably also a critical issue in all these large compilations. Another possibility is to generate simulated open clusters for different groups to analyse. A comparable `blind' analysis was initiated by \citet{Lebz12} to explore the reliability of different approaches to derive the effective temperature, $\log g$, and elemental abundances of cool giants. 

Next, we compared the cluster parameter surveys with an external sample for which we used mean parameter values derived from a large compilation of individual studies. These calibrators can be considered  a sample of true open clusters, which are confirmed by at least two individual studies. We derived correction terms by excluding very young and very old objects because both groups showed larger systematic deviations for various reasons. We found that most survey results are in general properly scaled, however, some show trends or constant offsets of varying degree. One study (T08) was excluded because the  results were too erroneous. This is probably because of their isochrone-fitting approach, which is   oriented too much to the brightest objects. For another work (G10), which also includes a large percentage of little studied objects, we were only able to verify the parameters by a comparison with the corrected results by K13. Thus, the derived corrections are probably biased by the sample selection. The surveys show a comparable accuracy after the correction, the mean intrinsic error (1$\sigma$) amounts to about 0.2\,dex for the age, 0.08\,mag for the reddening, and 0.35\,mag for the distance modulus. These values are clearly more promising than the errors given earlier. We noticed that the visual studies have an  accuracy that is twice as good for the reddening than the 2MASS surveys. Thus, the reddening is probably more difficult to derive in the NIR. The inclusion of visual data in the cluster analysis will certainly help to further improve the accuracy of all parameters. Furthermore, we found that an extinction ratio \ejk/\ebv\ = 0.46 provides the best match to the external calibration sample.

Finally, we found that the extensive 2MASS studies by B11 and K13, but probably also G10,
include at least 20\,\% of the problematic objects. For this percentage, the cluster results differ significantly for several reasons. These could be poorly populated objects with an ambiguous CMD, clusters with a high degree of field star contamination for which all cleaning procedures failed, or even doubtful or unlikely open clusters, which  do not facilitate an unambiguous fitting solution. The reason that parameters are almost always provided in the literature might be related to the challenging assumption about the reality of the cluster. Although, K13, for example, already excluded 11 percent  from their input list as
dubious objects. Nevertheless, the derived non-negligible percentage has to be taken into account if these parameter compilations are used for some statistical studies of the cluster population. Although most surveys provide reasonable results for the majority of objects, the blind use of the parameters of a single object, for example, to derive the luminosity function or to derive the mass for a particular cluster star, could lead to erroneous results as well. However, this will probably also hold for cluster parameters taken from individual studies.

\onlfig{
\begin{figure*}
\centering
\includegraphics[width=170mm]{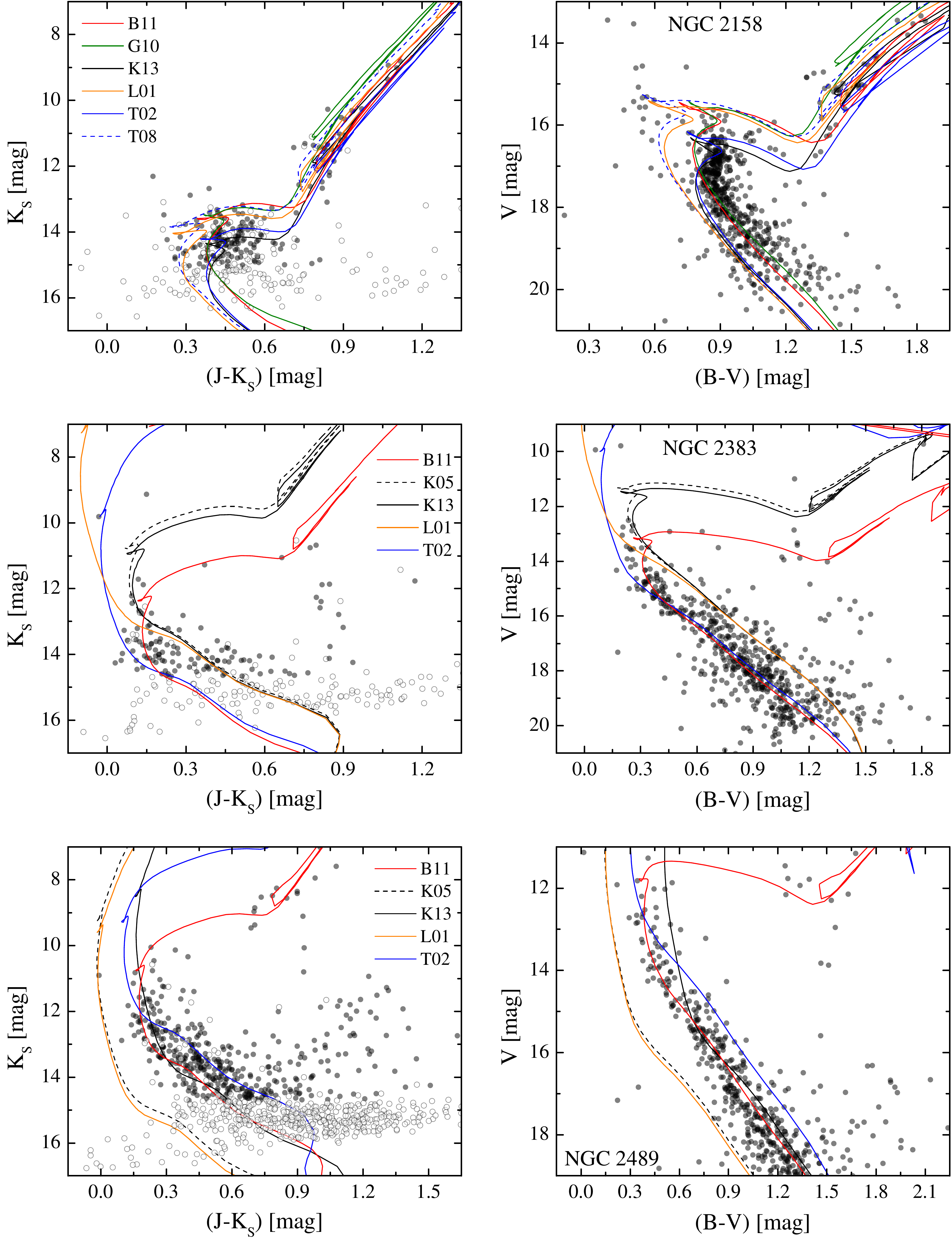}
\caption{The CMDs for the open clusters NGC~2158, NGC~2383, and NGC~2489. The \bv\ data were taken from \citet{carraro02}, \citet{subramaniam99}, and \citet{piatti07}, and we adopted a cluster radius of 1\farcm5, 2\farcm5, and 5\arcmin\ for NGC~2158, NGC~2383, and NGC~2489, respectively. The symbols are the same as in Fig. \ref{fig:ic4651}.} \label{cmd1}
\end{figure*}

\begin{figure*}
\centering
\includegraphics[width=170mm]{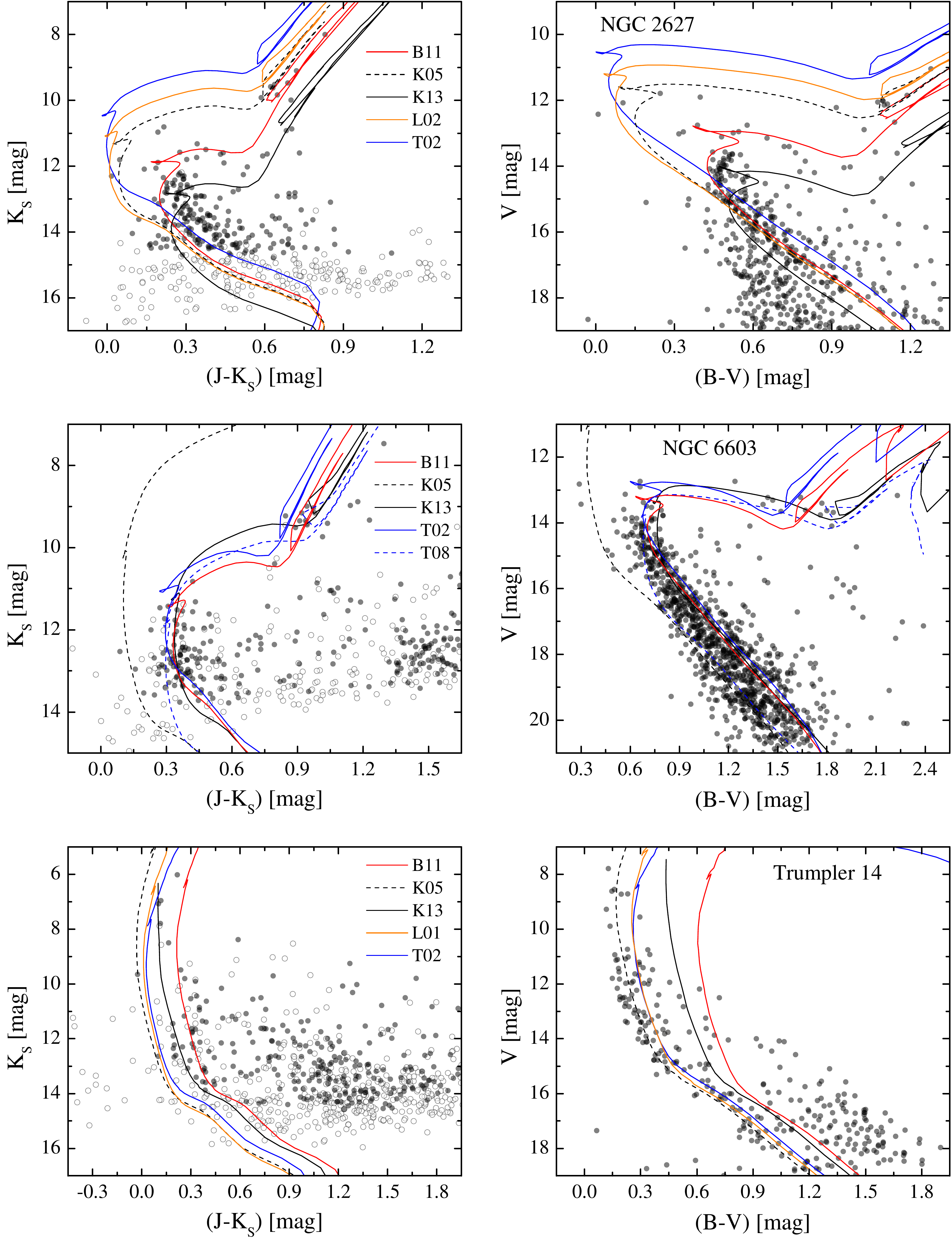}
\caption{The CMDs for the open clusters NGC~2627, NGC~6603, and Trumpler~14. The \bv\ data were taken from \citet{ahumada05}, \citet{sagar98}, and \citet{carraro04}, and we adopted a cluster radius of 4\arcmin, 2\arcmin, and 2\farcm5 for NGC~2627, NGC~6603, and Trumpler~14, respectively. The symbols are the same as in Fig. \ref{fig:ic4651}.} \label{cmd2}
\end{figure*}
}


\begin{acknowledgements}
This paper is dedicated to Anneliese Schnell who died during its preparation. We thank the anonymous referee for invaluable suggestions 
that have greatly improved the manuscript. MN acknowledges the support by the grant 14-26115P of the Czech Science Foundation GA\,\v{C}R. This project is also financed by the SoMoPro II programme (3SGA5916). The research leading
to these results acquired a financial grant from the People Programme
(Marie Curie action) of the Seventh Framework Programme of EU according to the REA Grant
Agreement No. 291782. The research is further co-financed by the South-Moravian Region. 
It was also supported by the grant 7AMB14AT015,
the financial contributions of the Austrian Agency for International 
Cooperation in Education and Research (BG-03/2013 and CZ-09/2014). This work reflects only the authors' views and the European 
Union is not liable for any use that may be made of the information contained therein.
This research has made use of the WEBDA database, operated at the Department of 
Theoretical Physics and Astrophysics of the Masaryk University. This publication also has made use of data products from the Two Micron All Sky Survey, which is a joint project of the University of Massachusetts and the Infrared Processing and Analysis Center/California Institute of Technology, funded by the National Aeronautics and Space Administration and the National Science Foundation, and was made possible through the use of the AAVSO Photometric All-Sky Survey (APASS), funded by the Robert Martin Ayers Sciences Fund.

\end{acknowledgements}



\bibliography{netopil_26372} 
\bibliographystyle{aa} 


\end{document}